# Dielectrophoretically Assembled Polymer Nanowires for Gas Sensing

Yaping Dan[a], Yanyan Cao[b], Tom E. Mallouk[b], Alan T. Johnson[a,c], Stephane Evoy[d]

[a]Department of Electrical and Systems Engineering, University of Pennsylvania, Philadelphia, PA 19104, USA

[b]Department of Chemistry, Pennsylvania State University, College Park, PA 16802, USA

[c]Department of Physics and Astronomy, University of Pennsylvania, Philadelphia, PA 19104, USA

[d]Department of Electrical and Computer Engineering and National Institute for Nanotechnology, University of Alberta, Edmonton, AB T6G 2V4, Canada

Abstract

We measured the electronic properties and gas sensing response of nanowires containing segments of poly(3,4-ethylenedioxythiophene)/poly(styrenesulfonate) (PEDOT/PSS) that were synthesized using anodic aluminum oxide (AAO) membranes. The nanowires have a "striped" structure of gold-PEDOT/PSS-gold and are typically 8 μm long (1 μm-6 μm-1 μm for each section, respectively) and 220 nm in diameter. Dielectrophoretic assembly was used to position single nanowires on pre-fabricated gold electrodes. A polymer conductivity of 11.5 ± 0.7 S/cm and a contact resistance of 27.6 ± 4 kΩ were inferred from resistance measurements of nanowires of varying length and diameter. When used as gas sensors, the wires showed a resistance change of 10.5%, 9%, and 4% at the saturation vapor pressure of acetone, methanol and ethanol, respectively. Sensor response and recovery were rapid (seconds) with excellent reproducibility in time and across devices. "Striped" template-grown nanowires are thus intriguing candidates for use in electronic nose vapor sensing systems.

Keywords: polymer nanowire vapor sensor dielectrophoretic assembly

# 1. Introduction

The development of a low-power and low-footprint integrated sensing array technology will open vast markets spanning a wide variety of applications such as clinical assaying, emission control, explosive detection, agricultural storage and shipping, and workplace hazard monitoring. Clinical diagnostic applications alone represent a significant market that is already driving significant efforts. For example, the acetone concentration in the breath of a healthy person is around 5 ppm, but this increases to 300 ppm for a patient with diabetes mellitus [1]. Mercaptans and aliphatic acids are found in the breath of patients with liver cirrhosis, while dimethyl- and trimethylamine are present in the breath of uremic patients [2].

Chemical sensors based on nanoscale materials offer exceptional opportunities to study the scaling of relevant physical and chemical phenomena, and present significant advantages over macroscopic sensors. It has been proposed that one-dimensional "nanowire" sensing elements possess significantly increased sensitivity over thin films sensors due to the one-dimensional confinement of carriers in such structures and their high surface-to-volume ratio [3]. In addition, dielectrophoretic assembly can be exploited to integrate large arrays of nanowire-based sensors onto silicon chips where signal processing circuitry has been fabricated using standard CMOS processes [4,5].

Conducting polymers are intriguing candidates for vapor sensing applications as they respond selectively to volatile organic compounds with high sensitivity under ambient conditions [6,7,8,9]. Poly(3,4-ethylenedioxythiophene)/poly(styrenesulfonate) (PEDOT/PSS) received sustained interest in recent years as one of the most stable conducting polymers [10].

Here we report on the electronic properties and gas sensing responses of PEDOT/PSS nanowires (~8 μm in length and ~220 nm in diameter) grown using a nanoporous template.

**2. Experimental**

In order to establish effective contacts with gold electrodes, the nanowires were synthesized with a "striped" structure of Au-PEDOT/PSS-Au using anodic aluminum oxide (AAO) membrane templates by slight modification of a published method [11,12]. Briefly, a thin layer of Ag was first evaporated onto one side of the AAO membrane to serve as the working electrode. The first Au segment of the nanowires was then deposited galvanostatically at 0.65 mA/cm$^2$ current density using a commercial Au plating solution (Orotemp 24 from Technic Inc.). The polymer segment was grown in a solution of 50 mM monomer and 100 mM poly(styrenesulfonic acid) (PSS) in 1:1 water/acetonitrile at a constant potential of 1.0 V. The choice of PSS as the dopant appeared to be important in preventing Au from entering the polymer matrix during the following electroless deposition, presumably because it prevented exchange of anions into the film from the electroless plating solution. The second Au segment was grown using electroless deposition [13] and then elongated by electrodeposition galvanostatically at 0.65 mA/cm$^2$ current density after the membrane was successively immersed into 25 mM $SnCl_2$ and 70 mM trifluoroacetic acid in 50:50 methanol/water (3 min), 30 mM $AgNO_3$ in 300 mM aqueous ammonia (2 min), and a solution containing 8 mM $Na_3Au(SO_3)_2$/129 mM $Na_2SO_3$/625 mM HCHO adjusted to pH 7 (over 24 h). After each solution, the membrane was rinsed thoroughly with deionized water.

The nanowires were released by dissolving the membrane in 500 mM NaOH for ~ 1 h, and the solution was then purified by repeating the following steps 5 times: centrifuge (using Marathon 16000R) at 1000RCF(=1000g) for 5 min, keep the bottom part of the solution and dilute it with deionized water. The nanowires were assembled from the solution onto specific locations using a home-built dielectrophoretic assembly apparatus. SEM images of samples after assembly clearly show that nanowires with the desired "striped" geometry have been formed through the synthetic process (Fig. 3, inset).

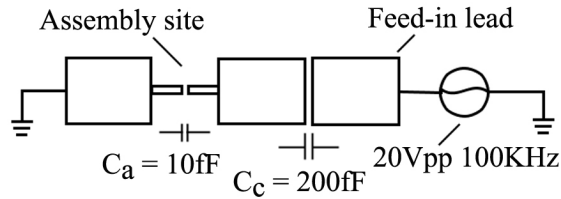

Fig. 1 Schematic of the on-chip circuitry used for dielectrophoretic assembly. The value of the coupling capacitor $C_c$ is designed to be roughly 20 times larger than that of the assembly capacitor $C_a$.

The device assembly site (Fig. 1) consisted of a pair of electrodes separated by approximately 6 μm (the assembly capacitor). One of the two electrodes was capacitively coupled by an 8 μm gap (the coupling capacitor) to a feed-in lead, and the second electrode was grounded. The coupling capacitor and assembly capacitor were designed to be approximately 200 fF and 10 fF, respectively. A voltage of 20 V peak-to-peak at a frequency of 100 kHz was applied across these two capacitors whose impedances at this frequency (8 MΩ and 150 MΩ, respectively) are ~100 - 1000 times greater than the polymer nanowires (~100 kΩ). These impedance values imply that before assembly about 19 V drops across the assembly capacitor, leading to a high electric field in this gap that draws nanowires into the desired location. After the first nanowire is brought into contact with both electrodes, the voltage drop across the assembly capacitor decreases to ~200 mV, so the assembly process is self-terminating. With samples consisting of an array of 30 such electrode pairs, we were able to assemble single-nanowire circuitry with a yield of 35 – 50 %. After assembly, a small drop of silver paste was applied to short the coupling capacitor and enable dc electrical measurements.

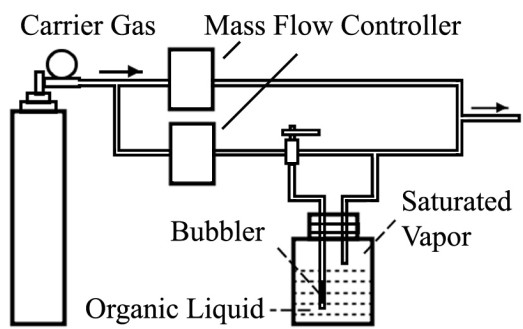

Fig. 2 The bubbler system used to deliver vapors of known analyte concentration to nanowire sensors.

Measurements of the electronic properties of nanowire devices were conducted under ambient conditions using a homebuilt apparatus. Multiple nanowires devices were measured simultaneously through the use of a solid state multiplexer (MAXIM DG406DJ) to switch electrical connections from nanowire to nanowire every 100 ms. Measurements of gas sensing responses were performed by enclosing nanowire devices in a chamber (1 × 1 × 1.5 cm), and exposing them to gas flows containing analyte vapors of known concentration created using a bubbler system (Fig. 2). Mass flow controllers regulated the flow of nitrogen (99.9% purity) carrier gas through two lines. One line was bubbled through the liquid analyte, providing a known flow rate of a saturated vapor. Saturated vapors ($P_0$) of the analytes used here are, in parts per thousand: methanol, 130; ethanol, 32.5; acetone, 520. The second nitrogen line was mixed with the flow of saturated vapor to bring the total gas flow to a standard level (typically 1000 sccm). By controlling the relative flow of the two lines, vapor of known analyte concentration ($P/P_0$) was created and then delivered to the sample while monitoring its electrical resistance. Pure nitrogen was used to flush the sample chamber by switching the two-way valve (Fig. 2).

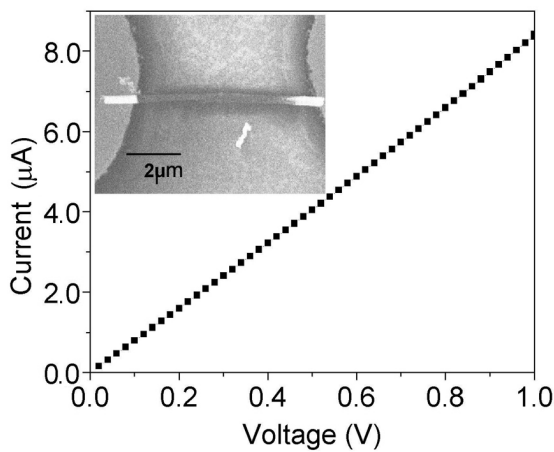

Fig. 3 Current-voltage characteristic of a single polymer nanowire. Inset: SEM image of a nanowire after dielectrophoretic assembly. The striped structure of the nanowire (bright gold ends, gray polymer center) is clearly visible.

## 3. Results and Discussion

The inset of Fig. 3 shows a single Au-PEDOT/PSS-Au nanowire assembled into position using dielectrophoresis. Devices typically had a linear current-voltage (I-V) characteristic (Fig. 3).

In order to determine $\sigma$, the electrical conductivity of the polymer, and the contact resistance $R_C$, the electronic properties of fourteen devices were measured. For each device the length and diameter of the polymer region of the nanowire were measured by AFM and/or SEM. With the assumption that the contact resistance $R_C$ is constant for each nanowire, the device resistance R is expected to be:

$$R = R_C + \frac{l}{\sigma A} \qquad (1)$$

where $l$ and $A$ are the length and cross-sectional area of the nanowire's polymer segment, respectively. For the samples tested, the polymer portion of the nanowire was 6 ± 0.5 μm long and 220 ± 25 nm in diameter.

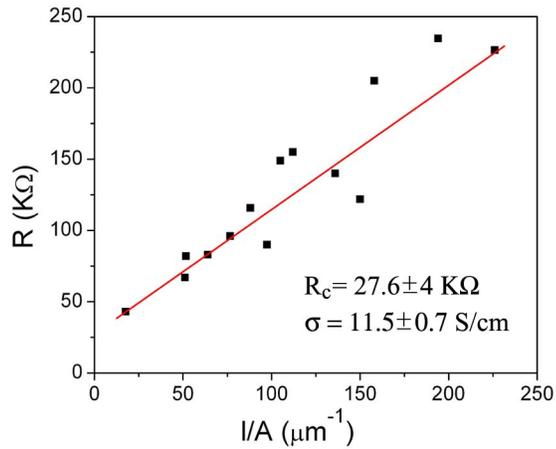

Fig. 4 Resistance vs. length/(cross-sectional area) $l/A$ is plotted for 14 samples.

Figure 4 shows a plot of R vs. $l/A$ for all samples, demonstrating that these quantities are linearly correlated as expected. The fit to Eq. 1 yields a contact resistance between PEDOT/PSS and gold of $R_C$ =27.6± 4 kΩ, and a polymer conductivity of σ = 11.5± 0.7 S/cm. This latter value is comparable to earlier reports [14,15] of the conductivities of thin films of similar polymers, which ranged from 0.03 to 80 S/cm, depending on the PEDOT:PSS ratio.

To characterize nanowire responses to analytes, we assembled devices using a new solution of nanowires from the same synthesis batch.

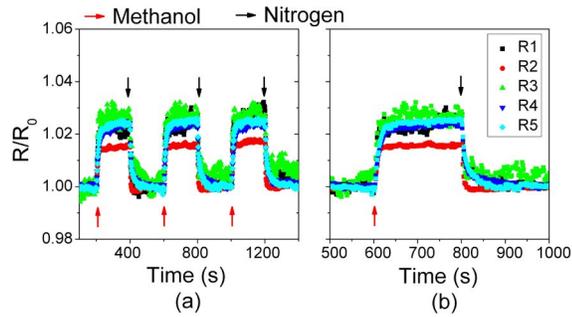

Fig. 5 (a) Relative resistance change of five nanowire devices simultaneously exposed to methanol at 20% of its saturated vapor pressure. (b) Enlargement of one exposure cycle.

Figure 5(a) shows the relative resistance $R/R_0$ (where $R_0$ is the nanowire resistance measured in a flow of pure nitrogen) of five nanowires (R1 to R5) upon exposure to 20% saturated vapor of methanol ($P/P_0 = 0.2$). The resistances $R_0$ of the 5 nanowires in pure nitrogen were 43.8 k$\Omega$, 32.8 k$\Omega$, 45.5 k$\Omega$, 42.2 k$\Omega$, and 48.4 k$\Omega$, respectively. Four of the nanowires showed indistinguishable responses of 2.2 ± 0.2%, while the fifth (R2) had a response of 1.5%. Although there was no obvious flaw in sample R2, it is clearly an outlier in both its initial resistance and methanol response. This observation of high sample-to-sample reproducibility in wire resistance and gas sensing response (with ~10% outliers) is typical of data from more than 40 nanowire devices. Because of its strong difference from the other devices, sample R2 was not included in measurements of the average gas sensing response presented below. Response to and recovery from analyte exposure were rapid, with all devices experiencing 90% of their resistance change within ~30 s.

The average response of four sensors to concentrations of methanol varying from 10-50% of a saturated vapor is plotted in Fig. 6(a). Sensor response (change of the normalized resistance $\Delta R/R_0$) increases linearly with methanol concentration as indicated in Fig. 6(b). Exposure to methanol concentrations larger than 50% of the saturated vapor pressure was found to lead to irreversible change in the nanowires resistance. Similar measurements were done for exposure to acetone and ethanol (Fig. 6b). The exposure to acetone vapor showed linear but larger response while that to ethanol vapor tended to saturate at high concentrations. Measurements of sensor response upon exposure to other analytes and at lower concentrations will be the subject of future work. Based on the current results, we expect to be able to detect these and other high vapor pressure analytes at concentrations well below 100 ppm.

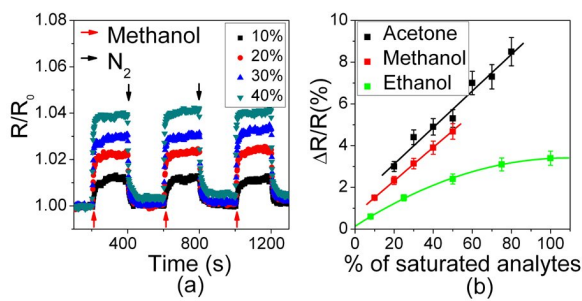

Fig. 6 (a) The average of the relative resistance of a set of four sensors increases with increasing methanol concentration. (b) Relative differential resistance response of the sensors as a function of analyte concentration for acetone, methanol, and ethanol. Lines are drawn as a guide to the eye.

We next discuss potential sensing mechanisms in this system and compare the results to previous reports. The observed increase in sensor resistance upon exposure to the analyte vapor may be due to polymer swelling [16,17] and/or a decrease in the charge carrier concentration on the polymer backbone of the doped PEDOT [18]. Strong evidence for irreversible swelling upon exposure to alcohol vapor, perhaps enhanced by substrate roughness, was observed in thin film sensors of PEDOT/PSS fabricated by inkjet printing [16]. Optical micrographs of the nanowire sensors used for this work showed no evidence of irreversible swelling on the micrometer scale (data not shown). However, we attribute the permanent damage to sensors exposed to methanol at 50% of the saturated vapor to a conformational change of the polymer chains that is not reversible.

The gas sensing properties of chemiresistor sensors based on PEDOT/PSS thin films [17,19] and random networks of nanorods of PEDOT doped with Fe-cations [18] have been previously reported. To our knowledge this is the first report of vapor sensors based on individual PEDOT nanowires. Thin film samples were seen to have markedly slower response times than our nanowire sensors, 2-3 min [17,18] compared to 30 s, while thick film sensors show response times exceeding 1 h [17]. Additionally, resistance changes of thin films exposed to methanol concentrations as low as 5 % of the saturated vapor are permanent [17], while the nanowire sensors show almost complete reversibility. The rapid response and recovery observed for nanowire sensors are consistent with the expectation of improved properties associated with their high surface-to-volume ratio. Moreover, in contrast to the nanowires sensors explored here, thin film sensors show significant baseline drift to the point of sensor destruction after prolonged exposure to methanol at $P/P_0 = 0.05$ [17].

In contrast to the superior response and recovery of the nanowire sensors, thin film samples had a higher sensitivity to methanol, defined as $(\Delta R/R_0)/(P/P_0)$, 0.2 – 2, compared to 0.15 for single nanowire sensors. The reverse was found in other experiments where sensors consisting of random networks of PEDOT nanorods doped with Fe-cations were nearly an order of magnitude more sensitive to $NH_3$ and HCl than thin films [18]. The origin of this discrepancy may lie in the impact of substrate roughness during film formation; the films discussed here were formed by inkjet printing on polyester film [17], spin coating onto glass slides [19], and oil-water interface polymerization [18]. In the case of inkjet printing, irreversible changes in film morphology were seen upon exposure to methanol at $P/P_0 = 0.05$ [17].

The PEDOT/PSS nanowire sensors do not show very high selectivity, as expected for this material system. However, they have a number of properties making them appropriate for use in "electronic nose" detectors, inspired by olfactory systems in biological organisms that typically utilize a thousand different odor receptors, each responsive to many different odorants, to perform amazing feats of molecular identification and analysis. PEDOT/PSS nanowires are all-electronic sensors that are reproducible in time and between sensors, with high sensitivity and fast response time (seconds). They offer the advantages of small footprint, simple implementation, and simple readout equipment, compared to chemicapacitors or sensors where vapor detection is done optically. Moreover, the approach described here, based on template-grown nanowires and dielectrophoretic assembly, may be extended to other polymeric, non-polymeric, or even composite materials with wide chemical diversity.

## 4. Conclusion

The conducting PEDOT/PSS nanowires described here are a very promising class of nanoscale gas sensors. They have good analyte sensitivity, rapid response and recovery, and excellent reproducibility, both in time and between samples, under ambient conditions. The "striped" structure of the nanowires (gold-PEDOT/PSS-gold) makes it straightforward to establish low resistance electrical contacts to gold leads using dielectrophoretic assembly, a method that is known to be compatible with CMOS microchips produced at conventional foundries. Moreover, it should be straightforward to extend the striped structure to other material systems to enable an electronic nose system. This system thus offers a promising method for constructing low-cost high-density integrated sensor arrays for a wide variety of gaseous analytes.

**Acknowledgment**

This work was supported by the National Science Foundation under NIRT grant ECS-0303981

**Biography of authors**

**Yaping Dan** received his bachelor and master degrees in microelectronics from Xi'an Jiaotong University in 1999 and Tsinghua University in 2002. He is currently a PhD student in the department of Electrical and Systems Engineering at the University of Pennsylvania. His research interests include nanoscale gas sensors, dielectrophoretic assembly, nanowire synthesis and nanofabrication.

**Yanyan Cao** received her bachelor's degree in Applied Chemistry and her M.S. degree in Supramolecular Chemistry from Tongji University. She is now a graduate student in Thomas E. Mallouk's group at Penn State, where she is working on the design, synthesis and application of functional materials especially for chemical sensing.

**Thomas E. Mallouk** was a graduate student at the University of California, Berkeley, and a postdoctoral fellow at MIT. In 1985, he joined the Chemistry faculty at the University of Texas at Austin. In 1993 he moved to Penn State, where he is now DuPont Professor of Materials Chemistry and Physics. His research has focused on the application of inorganic materials to different problems in solid state and surface chemistry, including photochemical energy conversion, nanoscale electronics, catalysis and electrocatalysis, chemical sensing, superconductivity, and environmental remediation.

**A.T. Charlie Johnson** was a graduate student at Harvard University and a postdoctoral fellow at the Delft University of Technology and the National Institute of Standards and Technology (Boulder). He has been at the University of Pennsylvania since 1994 and is now an Associate Professor in the Department of Physics and Astronomy, with secondary appointments in Electrical and Systems

Engineering, and Materials Science and Engineering. His current research interests include gas sensing using carbon nanotubes functionalized with biomolecules and polymer nanowires, single molecule electronics, and the synthesis of "programmable" nanostructures based on semiconductor nanowires and carbon nanotubes

**Stephane Evoy** received a Ph.D. in applied physics from Cornell University in 1998. He is Assistant Professor of Electrical and Computer Engineering at the University of Alberta, with cross-appointment as Leader of the Devices and Sensors Group of the National Institute for Nanotechnology. His current research includes the synthesis of nanomaterials of tunable properties, the development and integration of nanomechanical devices for biosensing applications, as well as the integration of nanostructures for the development of chemical sensing arrays.